\documentclass[%
 reprint,
superscriptaddress,
nofootinbib,
 amsmath,amssymb,
 aps,
]{revtex4-1}

\usepackage{graphicx}% Include figure files
\usepackage{dcolumn}% Align table columns on decimal point
\usepackage{bm}% bold math
\usepackage{diagbox}
\usepackage{tikz-feynman}
\usepackage{multirow}
\usepackage{braket}
\newcommand{\SUN}{{SU(N)}}
\newcommand{\Tr}[1]{{\rm Tr}\left[#1\right]}

\newcommand{\ie}{\textit{i.e.}~}
\newcommand{\diff}{\mathrm{d}}
\newcommand{\Diff}{\mathrm{D}}
\renewcommand{\epsilon}{\varepsilon}

\DeclareMathOperator{\SU}{SU}
\newcommand{\T}{\mathbb{T}}
\renewcommand{\tilde}{\widetilde}

\DeclareMathOperator{\tr}{tr}

\newcommand{\Bbar}{\bar{B}}
\newcommand{\mbarsq}{\overline{m}^2}
\makeatletter
\def\p@paragraph{\thesection\thesubsection}% ...removed \thesubsubsection\,
\makeatother
\begin{document}

\preprint{APS/123-QED}

\title{Nonperturbative infrared finiteness \\in super-renormalisable scalar quantum field theory}% Force line breaks with \\
%\thanks{}%

\author{Guido Cossu}
\affiliation{Braid Technologies, Shibuya 2-24-12, Tokyo, Japan}
\affiliation{Higgs Centre for Theoretical Physics, School of Physics and Astronomy, The University of Edinburgh, Edinburgh EH9 3FD, UK}
\author{Luigi~Del~Debbio}
\affiliation{Higgs Centre for Theoretical Physics, School of Physics and Astronomy, The University of Edinburgh, Edinburgh EH9 3FD, UK}%Lines break automatically or can be
\author{Andreas~J\"uttner}
\email{juettner@soton.ac.uk}
\affiliation{School of Physics and Astronomy, University of Southampton, Southampton SO17 1BJ, UK}
 \affiliation{STAG Research Center,
University of Southampton, Highfield, Southampton SO17 1BJ, UK}
\author{Ben~Kitching-Morley}
\affiliation{School of Physics and Astronomy, University of Southampton, Southampton SO17 1BJ, UK}
\affiliation{Mathematical Sciences, University of Southampton, Highfield, Southampton SO17 1BJ, UK}
\affiliation{STAG Research Center, University of Southampton, Highfield, Southampton SO17 1BJ, UK}
\author{Joseph~K.~L.~Lee}%
\author{Antonin~Portelli}%
\affiliation{Higgs Centre for Theoretical Physics, School of Physics and Astronomy, The University of Edinburgh, Edinburgh EH9 3FD, UK}%Lines break automatically or can be
\author{Henrique~Bergallo Rocha}
\affiliation{Higgs Centre for Theoretical Physics, School of Physics and Astronomy, The University of Edinburgh, Edinburgh EH9 3FD, UK}%Lines break automatically or can be
\author{Kostas~Skenderis}
 \affiliation{Mathematical Sciences,
University of Southampton, Highfield, Southampton SO17 1BJ, UK}
\affiliation{STAG Research Center, University of Southampton, Highfield, Southampton SO17 1BJ, UK}

 \collaboration{LatCos Collaboration}

\date{\today}% It is always \today, today,

\begin{abstract}
We present a study of the IR behaviour of a three-dimensional super-renormalisable  quantum field theory (QFT) consisting of a scalar field in the adjoint of $SU(N)$ with a $\varphi^4$ interaction.
A bare mass is required for the theory to be massless at the quantum level. In perturbation theory the critical mass is ambiguous due to infrared (IR) divergences and we indeed find that at two-loops in lattice perturbation theory the critical mass diverges logarithmically. It was conjectured long ago in \cite{Jackiw:1980kv,Appelquist:1981vg}  that super-renormalisable theories are nonperturbatively IR finite, with the coupling constant playing the role of an IR regulator. Using a combination of Markov-Chain-Monte-Carlo simulations of the lattice-regularised theory, both frequentist and Bayesian data analysis, and considerations of a corresponding effective theory we gather evidence that this is indeed the case.

\end{abstract}

 \maketitle

\paragraph{Introduction:}
Massless super-re\-nor\-ma\-lis\-able quan\-tum field theories suffer from severe infrared (IR)
divergences in perturbation theory: the same power counting argument 
that implies good ultraviolet (UV) behavior also implies bad IR behavior. Explicit
perturbative computations (with an IR regulator) lead to IR logarithms which make the 
perturbative results ambiguous. The fate 
of the IR singularities was discussed in \cite{Jackiw:1980kv,Appelquist:1981vg} where 
it was argued that such theories 
are nonperturbatively IR finite. In the examples analysed in \cite{Jackiw:1980kv,Appelquist:1981vg} the nonperturbative answer, when 
expanded with a small coupling constant, reduced to the perturbative result but with the IR regulator replaced by 
the (dimensionful) coupling constant. 

One motivation for the original studies was that in the high-temperature limit of four-dimensional Yang-Mills (YM) theory there is an effective dimensional reduction to three dimensions and the dimensionally reduced 
theory is super-renormalizable~ (see, for example, \cite{Appelquist:1981vg,Farakos:1994kx,Kajantie:1995kf}). Here our motivation comes from a new application of massless super-renormalisable 
theories: such theories appear in holographic models for the very early universe~\cite{McFadden:2009fg}. 

The models introduced in~\cite{McFadden:2009fg} are based on a three-dimensional $SU(N)$ Yang-Mills theory coupled to massless scalars $\varphi$ in the adjoint of $SU(N)$ with a $\varphi^4$ interaction. 
To compute the predictions of these models for cosmological observables one needs a nonperturbative evaluation of the relevant QFT correlators. This is the case even in the regime where the effective expansion parameter is small
because of the IR singularities discussed above. Moreover, understanding the IR behavior of this QFT is important for another reason: in holographic cosmology cosmic evolution corresponds to inverse RG flow, and the initial singularity in the bulk is mapped to the IR behavior of the dual QFT. Thus a mechanism for curing the IR singularities would also provide a holographic resolution of the initial bulk singularity. 

In this Letter we initiate the study of such a theory using lattice methods. We will focus on the simplest theory within this class: three-dimensional massless scalar QFT with $\varphi$ in the adjoint of $SU(N)$ and a $\varphi^4$ interaction regularised on a Euclidean space-time lattice~\cite{Lee:2019zml,DelDebbio:2020amx}. 
It turns out this theory still provides an interesting holographic model.
Irrespective of the holographic motivation we believe that understanding the fate of IR singularities in this QFT is an interesting problem in its own right and this model provides the possibility to explicitly test the hypothesis in~\cite{Jackiw:1980kv, Appelquist:1981vg}.

We address two central questions in this paper: Is the theory nonperturbatively IR finite, and what is the critical mass, {\it i.e.} what is the value of the bare mass such that the renormalised theory is massless? The latter question 
is crucial for future simulations at the massless limit where the holographic duality is defined~\cite{McFadden:2009fg}.
{Through two-loops} the critical mass is
both  linearly UV divergent and logarithmically IR divergent. We proceed to a nonperturbative
determination of the  critical mass in Markov-Chain-Monte-Carlo simulations of the discretised 
Euclidean path integral, where naively  the inverse of the finite extent of the lattice $L$ acts as the only IR regulator. 
By studying the finite-size scaling (FSS) nonperturbatively, within the effective theory and on the lattice,
we find evidence for the absence of the IR divergence beyond perturbation theory. 

The $N=2$ model is equivalent to the $O(3)$ vector model and the $N=3$ model is in the same universality class as the $O(8)$ vector model~\cite{Delfino:2015}, which have been studied widely in the literature \cite{Campostrini:2002ky}, including studies of their critical mass and other values of $N$~\cite{Pelissetto:2015yha,Arnold:2001ir,Sun:2002cc}. For $N>3$ such an equivalence is not obvious and little is known about the theories' phase
structure (see, for example, \cite{Pelissetto:2019zvh}). %\cite{Arnold:2001ir,Delfino:2015}.

\paragraph{Lattice perturbation theory:}
We consider the three-dimensional Euclidean scalar $\SUN$ valued $\varphi^4$ theory, 
\begin{equation}\label{eq:canonical S}
S=\int d^3x\,\Tr{\left(\partial_\mu\varphi(x)\right)^2+(m^2-m_c^2)\varphi(x)^2+\lambda\varphi(x)^4}\,,
\end{equation}
with fields $\varphi=\varphi^a(x) T^a$ where $\varphi^a(x)$ is real, and 
$T^a$ are the generators of $\SUN$ ($\Tr{T^aT^b}=\frac 12 \delta_{ab}$). 
In the following we prefer to work with a rescaled version of the action where the 't Hooft scaling is explicit,
\begin{equation}
S=\frac{N}g\int d^3x\,\Tr{\left(\partial_\mu\phi(x)\right)^2+(m^2-m_c^2)\phi(x)^2+\phi(x)^4}\,,
\end{equation}
which we obtain from Eq.~(\ref{eq:canonical S}) by identifying
$\phi=\sqrt{N/g}\,\varphi$ and $\lambda=g/N$, where $g$ is the 't Hooft coupling which should be kept fixed in  the large $N$ limit. 
The field $\phi$ and coupling constant $g$ have mass dimension one.

To discretise the theory on a $3d$ space-time lattice with lattice spacing $a$ we replace
partial derivatives by finite differences,
$\partial_\mu \phi(x)\to\delta_\mu\phi(x)=\left(\phi(x+\hat\mu a)-\phi(x)\right)/a$,
and integrals by sums
$\int d^3x \to a^3\sum_{x\in\Lambda^3}$,
where $a$ is the lattice constant, $\hat\mu$ a unit-vector in the $\mu$ direction
and $\Lambda^3$ the set of all lattice sites. We use 
periodic boundary conditions.

\begin{figure}
    \begin{center}
    \includegraphics[width=8cm]{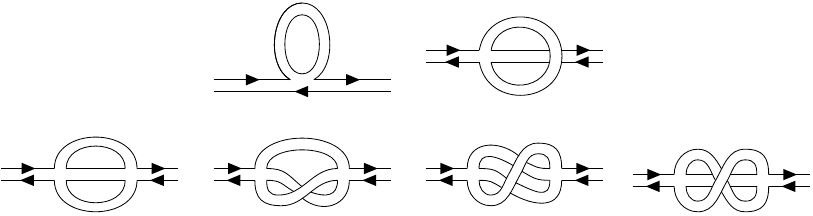}
    \end{center}
    \caption{One- and two-loop diagrams contributing to the mass-renormalisation in double-line representation representing matrix indices of the 
    scalar propagator.}\label{fig:2loop diagrams}
\end{figure}
The diagrams that contribute to the critical mass, $m_c^2$, at the two-loop level are shown in Fig.~\ref{fig:2loop diagrams}. The 
IR-finite but linearly UV-divergent one-loop integral  is
\begin{equation}\label{eq:I0}
 \int \limits_{-\pi/a}^{\pi/a}\frac {d^3k}{(2\pi)^3}\frac1{\hat k^2}=\frac {Z_0}{a}\,\, {\rm with}\,\,Z_0=0.252731...\,,
\end{equation}
for lattice momenta $\hat k=\frac 2a \sin(ka/2)$. The integral to
be evaluated at two-loop with vanishing external momentum $p=0$ is
\begin{equation}\label{eq:2loopI}
D(p)=\int\limits_{-\pi/a}^{\pi/a}\frac {d^3k}{(2\pi)^3}\frac{d^3q}{(2\pi)^3}\frac 1{\hat k^2\,\hat q^2\, \hat r^2}\,,
\end{equation}
where $r=-k-q-p$, and \emph{hatted}
quantities are defined as above. 
By naive dimensional counting, and confirmed by repeating the analysis of the IR-properties of this diagram in \cite{Luscher:1995zz} for $d=3$, we
find that the integral diverges logarithmically in the IR,
\begin{align}\label{eq:IR analytical}
D(p)\stackrel{p\to 0}{=}D_{\rm IR}(p)=-\frac{\log(|pa|)}{(4\pi)^2}
\end{align}
(derivation in Sec.~\ref{sec:IR2loop} of supplementary material). 

Following~\cite{Jackiw:1980kv,Appelquist:1981vg}, we  choose to introduce an IR regulator by setting the external momentum to $g/(4\pi N)\equiv\Lambda$.\footnote{While the expression for the logarithmic cut-off dependence for a given UV regulator can be computed in perturbation theory~\cite{Laine:1995np,Laine:1997dy}, the precise form for the IR regulator is unknown.}  The two-loop expression for the critical mass then evaluates to 
\begin{equation}\label{eq:PTmc}
m_c^2(g)=-g\frac {Z_0}{a} \left(2-\frac 3{N^2}\right)+g^2 
{D}(\Lambda)\mathcal{N}(N)\,,
\end{equation}
where $\mathcal{N}(N)=1- 6/{N^2}+{18}/{N^4}$ (see also~\cite{Laine:1997dy}).
Representative values for $D(\Lambda)$ and $m_c^2(g)$ at
one- and two-loops for $N=2$ are listed in table Tab.~\ref{tab:PTres}.\footnote{We evaluate the two-loop lattice
momentum integral using the Markov Chain Monte Carlo integration implemented in VEGAS~\cite{VEGAS}. The error estimates we provide together with the results
 are statistical only.}
For the range of couplings presented in the table the change from one- to two-loop 
corresponds to a relative change in the range {1\% to 6\%}.
{Note that the $n$-loop ($n >2$) contribution to the critical mass is power-law IR-divergent, 
$\sim g^2 (g/\Lambda)^{n-2}$. If $\Lambda \sim g$, as in the mechanism of~\cite{Jackiw:1980kv,Appelquist:1981vg}, such terms are finite and proportional to $g^2$. On the other hand, if the theory is truly IR divergent such terms would dominate in the IR.}
\begin{table}
\begin{tabular}{l|l|ll}
\hline\hline
\multicolumn{1}{c|}{\multirow{2}{*}{$ag$}}&
\multicolumn{1}{c|}{\multirow{2}{*}{$D(\Lambda)$}}&
\multicolumn{2}{c}{$(am_c)^2$, $N=2$}\\%&
&&\multicolumn{1}{c}{1-loop}&\multicolumn{1}{c}{2-loop}\\
\hline
0.1 & 0.05469(19) & -0.03159 & -0.03125  \\%& -0.04581 & -0.04543 \\
0.2 & 0.04953(13) & -0.06318 & -0.06194  \\%& -0.09161 & -0.09024 \\
0.3 & 0.04783(13) & -0.09477 & -0.09208  \\%& -0.13742 & -0.13443\\
0.5 & 0.045311(92) & -0.15796 & -0.15088 \\%& -0.22904 & -0.22116\\
0.6 & 0.044134(90) & -0.18955 & -0.17962 \\%& -0.27484 & -0.26380\\
\hline\hline
\end{tabular}
\caption{Results for the two-loop integral $D(\Lambda)$ and the  critical mass in lattice perturbation theory.}\label{tab:PTres}
\end{table}

\paragraph{Finite-size scaling for $m_c^2$:}
In this section we provide details and results of our nonperturbative studies towards the 
determination of the critical mass. 
Our strategy is to compute it as a function of the IR cutoff given in terms of the inverse lattice size $1/L$, by means of FSS. The observable we consider is the 
Binder cumulant,
\begin{equation}\label{eq:B}
B = 1-\frac N3 \frac{\langle \Tr{M^4}\rangle}{\langle \Tr{M^2}\rangle^2}\,,
\end{equation}
where $M$ is the magnetisation matrix defined below, and $\langle \cdot\rangle$ indicates expectation value under the Euclidean path integral. 

For each choice of simulation parameters, we determine the bare input mass, 
$\mbarsq(\Bbar, g,L)$, in the critical region 
for which  the Binder cumulant takes some suitably chosen value  $\Bbar$.
The Binder cumulant in a finite volume of extent $L$ in the critical scaling region is described by a scaling function $f$,
\begin{align}\label{eq:FSSansatz}
\Bbar=f\left(\left(\mbarsq(\Bbar,g,L)-m_c^2(g)\right)/g^2\,x^{1/\nu}\right)\,, 
\end{align}
where $x=gL$ and $\nu$ is the critical exponent.
 Expanding $f$ in the vicinity of the critical mass we find the leading FSS behaviour

\begin{align}
\overline{m}^2(\Bbar,g,L)=&m_c^2(g)+\,g^2x^{-{1}/{\nu}}\frac {\Bbar 
-f(0)}{f'(0)}\,.\label{eq:mc naive scaling}
\end{align}

\paragraph{FSS in the Continuum Effective Theory:}\label{sec:EFT}
Before analysing and interpreting simulation data for the FSS 
of the critical mass, we can gain
further analytical understanding of the critical behaviour. 
To this end we consider the effective field theory (EFT) of the zero-mode of the field $\phi$, i.e. the magnetisation 
\begin{align}\label{eq:mag}
M=\frac 1{L^3}\int d^3x\,\phi(x)\,,
\end{align}
and fluctuations $\chi$ around it, i.e. $\phi=M+\chi$. In the vicinity of the critical point 
long{-}distance contributions
described by $M$ dominate, motivating us to consider the leading-order effective action
\begin{align}
S_{\rm eff}=\frac {L^3N}g\left[(m^2-m_c^2) \Tr{M^2}+\Tr{M^4}\right]\,.
\end{align}
Following~\cite{Zinn-Justin:572813}, we quantise the theory under the finite-volume 
path integral and find integral expressions for the Binder cumulant (for details 
see Sec~\ref{sec:treeeft} of supplementary material). 
Expanding
again in the vicinity of the critical point we recover Eq.~(\ref{eq:mc naive scaling})
and compute the leading-order predictions
 $\nu|_{N=2,4}=2/3$, $ f(0)|_{N=2}\approx 0.5431$ and $ f'(0)|_{N=2}\approx -0.03586$, and $ f(0)_{N=4}\approx 0.4459$ and $ f'(0)_{N=4}\approx -0.02707$, respectively.
%%%%%%%%%%%%%%%%%%%%%%%%%%%%%%%%%%%%%%%%%%%%%%%%%%%%%%%%%%%%
\paragraph{Lattice simulation:}
%%%%%%%%%%%%%%%%%%%%%%%%%%%%%%%%%%%%%%%%%%%%%%%%%%%%%%%%%%%%
We implemented the model in the GRID library~\cite{Boyle:2015tjk,Boyle:2016lbp} 
with both the Hybrid Monte Carlo~\cite{Duane:1987de} and a heat-bath over-relaxation 
algorithm~\cite{Brown1987b,Adler1988b,Fodor1994a,Bunk1995a}. We generated ensembles of $O(100{\rm k})$ field configurations 
for $N=2,4$,  volumes with $L/a=8,16,32,48,64,96,128$,   
couplings $ag=0.1,0.2,0.3,0.5,0.6$, and a number of bare mass parameters in the vicinity of the
perturbative prediction for $m_c^2(g)$ in Eq.~(\ref{eq:PTmc}). By using a wide range of couplings, a 
large range of lattice volumes was covered ($0.8\leq x\leq 76.8$) while keeping simulation
costs acceptable.\footnote{The simulation data as well as the Python
analysis code have been made publicly available as~\cite{kitching_morley_ben_2020_4290508,cossu_guido_2020_4266114}.}

\begin{figure}
\begin{center}
\includegraphics[width=8.5cm]{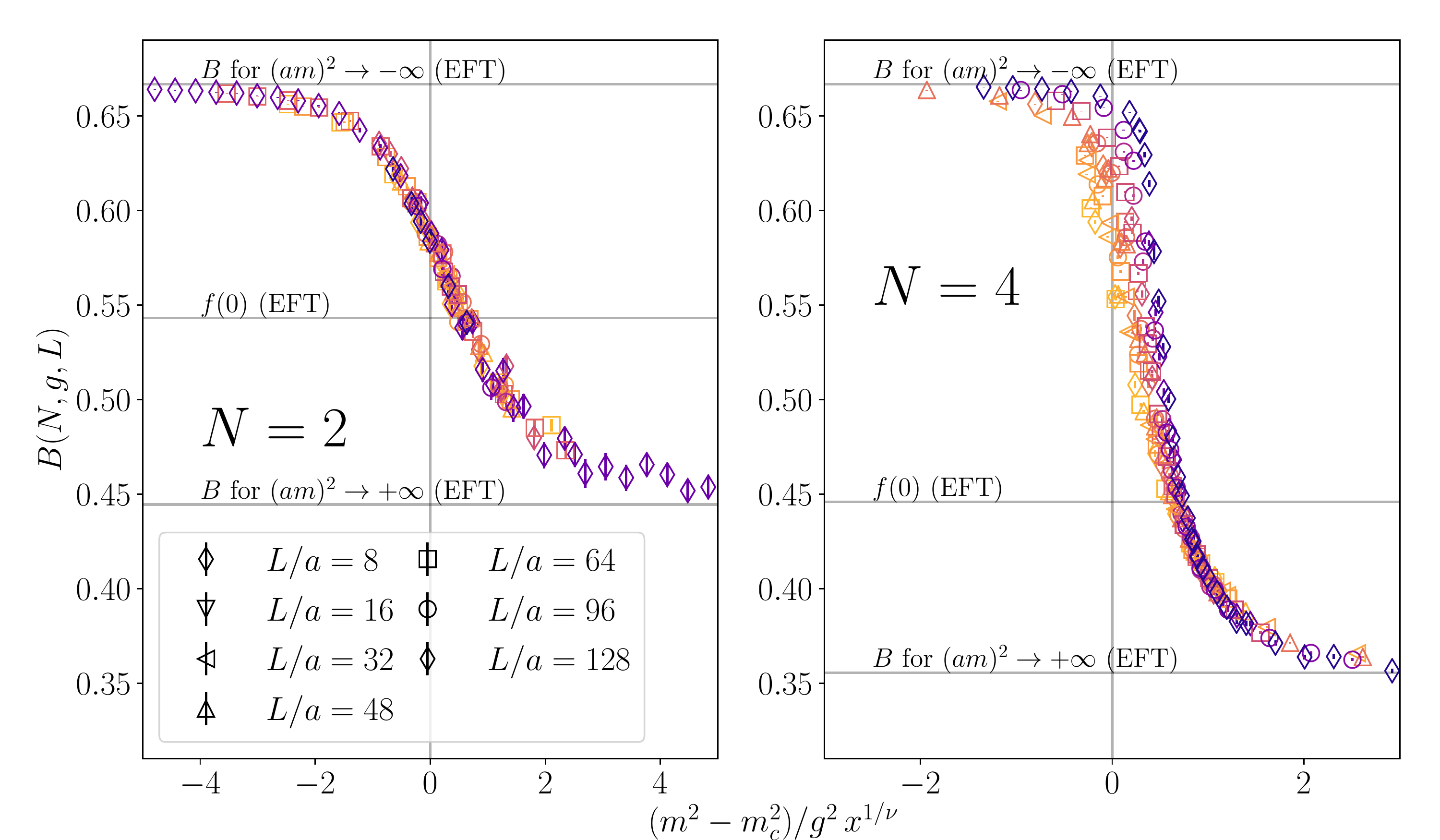}\\
\hspace{-3mm}\includegraphics[width=9cm]{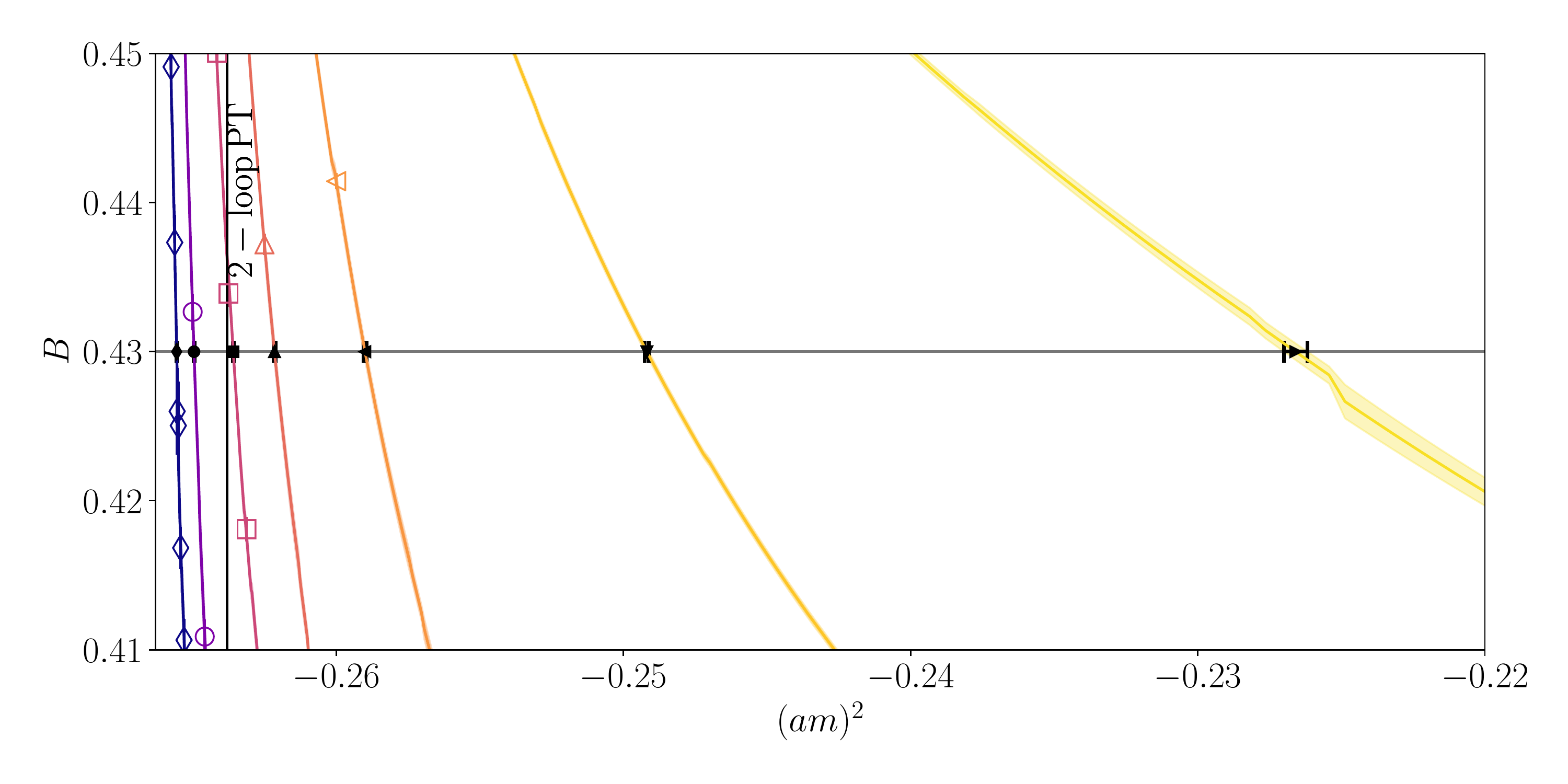}
\end{center}
\caption{Top: $N=2$ (left) and $N=4$ (right) results for the Binder cumulant, the EFT prediction for $f(0)$ and the value of the Binder cumulant in the limits $m^2\to\pm\infty$ (cf. Sec.~\ref{sec:EFT}). The values on the $x$-axis have been rescaled using the values of the critical exponent $\nu$ and the critical masses $m_c^2$ determined in Sec.~\ref{sec:FSS analysis}. Darker colour corresponds to larger value of $gL$. Bottom: Data points from simulations, lines from reweighting with width corresponding to the statistical error. Intersects of $N=4$, $g=0.6$ data for, from left to right, $L/a=128,96,64,48,32,8,16$ with $\Bbar =0.43$ indicated with $y$-error bars. 
The black vertical line
indicates the 2-loop infinite-volume value of
the critical mass.
}\label{fig:latticedata}
\end{figure}

Using multi-histogram reweighting~\cite{Ferrenberg:1988yz} we obtained a continuous 
representation of the Binder cumulant as a function of the bare scalar mass. Example results for $B(N,g,L)$ are shown in the top panel of Fig.~\ref{fig:latticedata} and
the reweighting is illustrated in
the bottom panel.
The analysis was carried out 
under bootstrap resampling~\cite{Efron}.
We determined the integrated autocorrelation time $\tau_{\rm int}$ for
$M^2$, $M^4$ and $\phi^2$ with the method of~\cite{Wolff:2003sm} with largest values being O(100).
All data was binned into bins of size $\rm max(50,4\tau_{\rm int})$.
The reweighting allows for a model-independent determination of 
${\overline m}^2(\bar B,g,L)$ by means of an iterative solution.
Example results for $\mbarsq(\Bbar,g,L)$ are listed in~ Tab.~\ref{tab:LatticeResults msq}. 
We note the proximity of these finite-volume results to the 2-loop infinite-volume prediction 
listed in Tab.~\ref{tab:PTres}.
\begin{table*}
\begin{tabular}{c|lllllllllllllllll}
\hline\hline
\diagbox{$L/a$}{$ag$}&\multicolumn{1}{c}{0.1}&\multicolumn{1}{c}{0.2}&\multicolumn{1}{c}{0.3}&\multicolumn{1}{c}{0.5}&\multicolumn{1}{c}{0.6}\\ 
\hline
8&-0.024289(87)  &-0.05048(16)  &-0.07717(13) &-0.12989(17) &-0.15680(31)\\
16&-0.028398(37)  &-0.057413(65) &-0.086163(97)&-0.143556(77)&-0.17205(13)\\
32&-0.030071(19)  &-0.060181(51) &-0.090135(40)&-0.149284(65)&-0.178777(53)\\
48&-0.030595(21)  &-0.061032(37) &-0.091267(47)&-0.151126(48)&-0.180582(51)\\
64&-0.030841(13)  &-0.061448(45) &-0.091814(26)&-0.151816(72)&-0.181522(28)\\
96&-0.031067(12)  &-0.061811(16) &-0.092270(31)&-0.152521(29)&-0.182345(66)\\
128&-0.0311266(93) &-0.061962(43) &-0.092486(29)&-0.152808(33)&-0.182680(29)\\
\hline\hline
\end{tabular}
\caption{Results for $(a\overline{m})^2(\Bbar=0.53,g,L)$ for $N=2$.}\label{tab:LatticeResults msq}
\end{table*}

%%%%%%%%%%%%%%%%%%%%%%%%%%%%%%%%%%%%%%%%%%%%%%%%%%%%%%%%%%%%
\paragraph{Finite{-}size scaling analysis:}\label{sec:FSS analysis}
%%%%%%%%%%%%%%%%%%%%%%%%%%%%%%%%%%%%%%%%%%%%%%%%%%%%%%%%%%%%
We now turn to the fitting of $\mbarsq(\Bbar,g,L)$. 
Guided by Eq.~(\ref{eq:mc naive scaling}) we chose the fit ansatz
\begin{align}\label{eq:msq fit}
\overline{m}^2(\bar B,&g,L)=m_c^2(g)|_{\rm 1-loop}+g^2\alpha
\nonumber\\
&+g^2\Bigg(x^{-{1}/{\nu}}\frac {
\bar B-f_0}{f_1}+\beta D_{\rm IR}(\Lambda_{\rm IR})\mathcal{N}(N)\Bigg)\,,
\end{align}
The first term is the 1-loop expression for the critical mass and it removes the linear UV divergence perturbatively (cf. Eq.~(\ref{eq:PTmc})). 
{The coefficient $\alpha$ includes potential residual scheme dependence in the IR/UV regulator, e.g. normalisation factors in the argument of $D_{\rm IR}$, as well as the contribution from higher loops when $\Lambda_{\rm IR}\sim g$.} 
The second term in brackets parameterises the dependence on the IR cutoff
for which we study, respectively, 
$\Lambda_{\rm IR}=\frac1{4\pi}\frac {g}{N} $ and $\frac 1{L}$. {In the case $\Lambda_{\rm IR}=\frac 1{L}$ the $n$-loop IR divergent contribution yields $g^2 x^{n-2}$, which is of the same form as the finite scaling correction but with effective scaling dimension that tends to zero as $n\to \infty.$ If such terms are present, their effects would dominate over the logarithmic or the finite size behavior in the IR.}  
 To better constrain the fit we simultaneously analyse data from 
 various pairings of two $\bar B$ values in the vicinity of $f(0)$ as predicted in the EFT~(cf. Sec.~\ref{sec:EFT}).  For $N=4$ we allowed
one value of $\alpha$ per $\Bbar$ value. For $N=2$ excellent fit quality was
achieved without this additional freedom.
 
\begin{figure}
\hspace{-.4cm}\includegraphics[width=9cm]{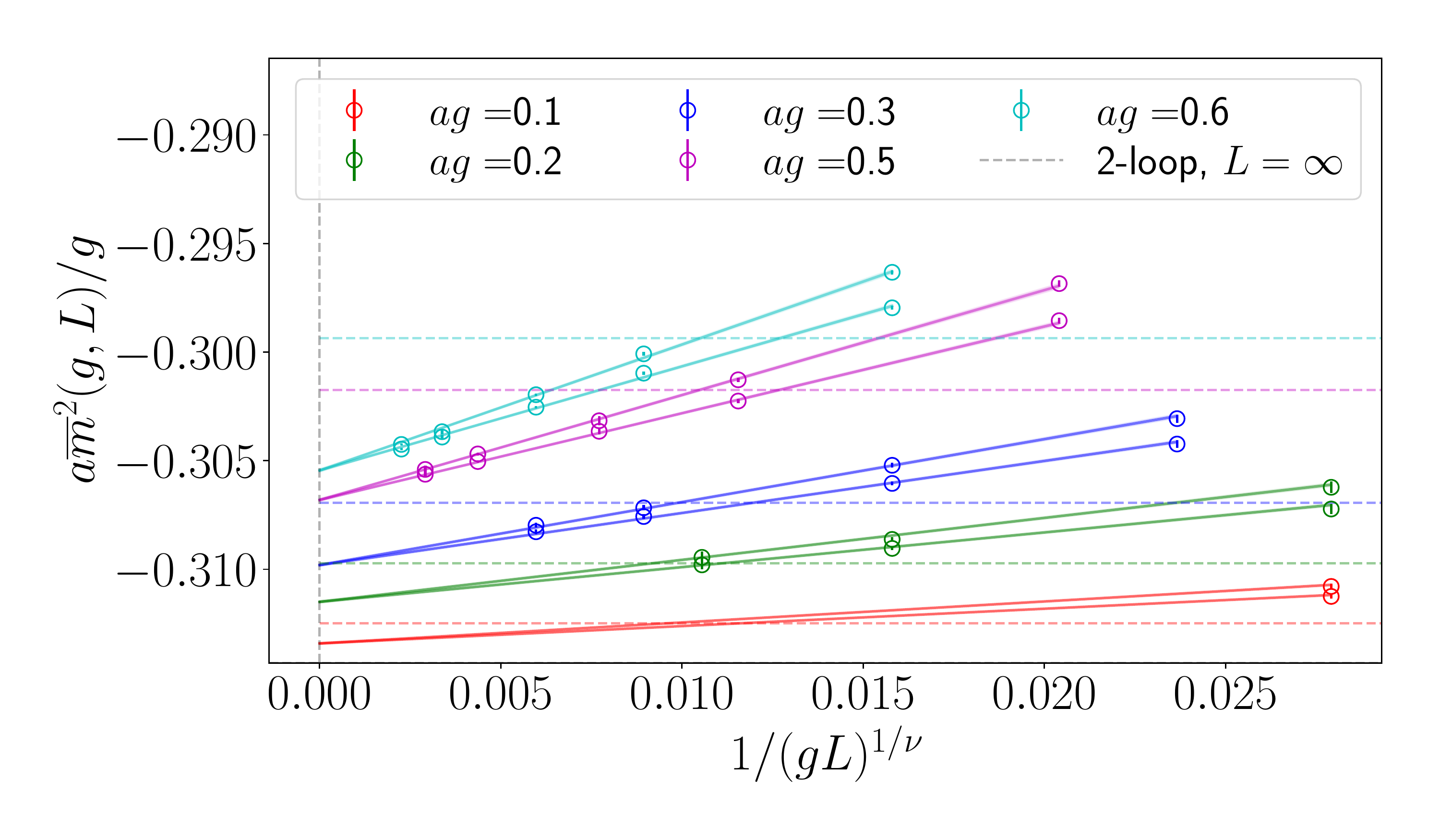}
\caption{Central fit $N=2$, $\Bbar=0.52,0.53$. Dashed lines correspond to the 2-loop prediction for the effective mass, solid lines to the fit result including error band. Value of $ag$ increasing from bottom to top. At each coupling the top set of points corresponds to $\Bbar=0.52$, bottom set to $\Bbar=0.53$.}\label{fig:centralfit}
\end{figure}

\begin{figure}
    \centering
    \includegraphics[width=80mm]{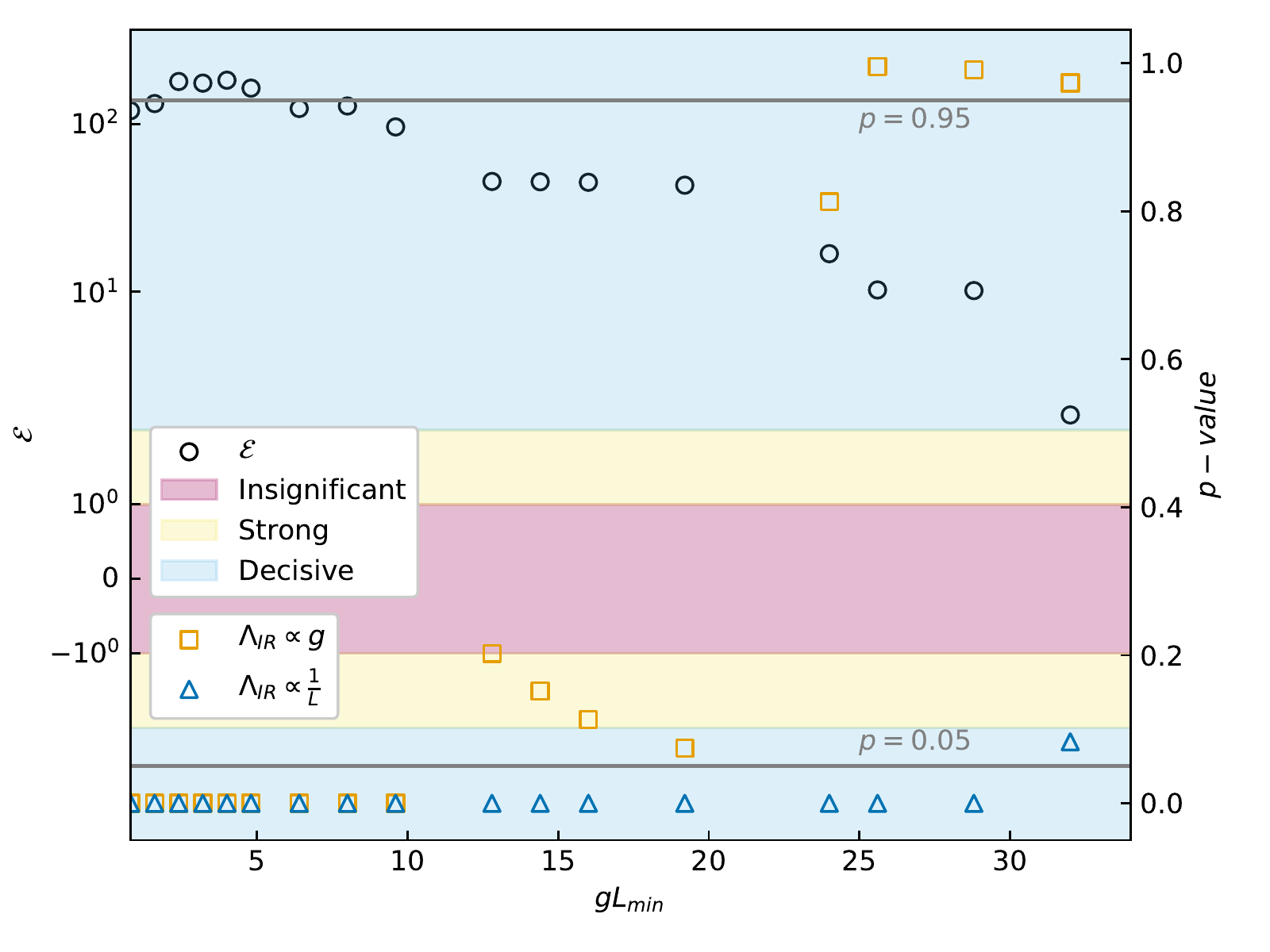}\\
    \includegraphics[width=80mm]{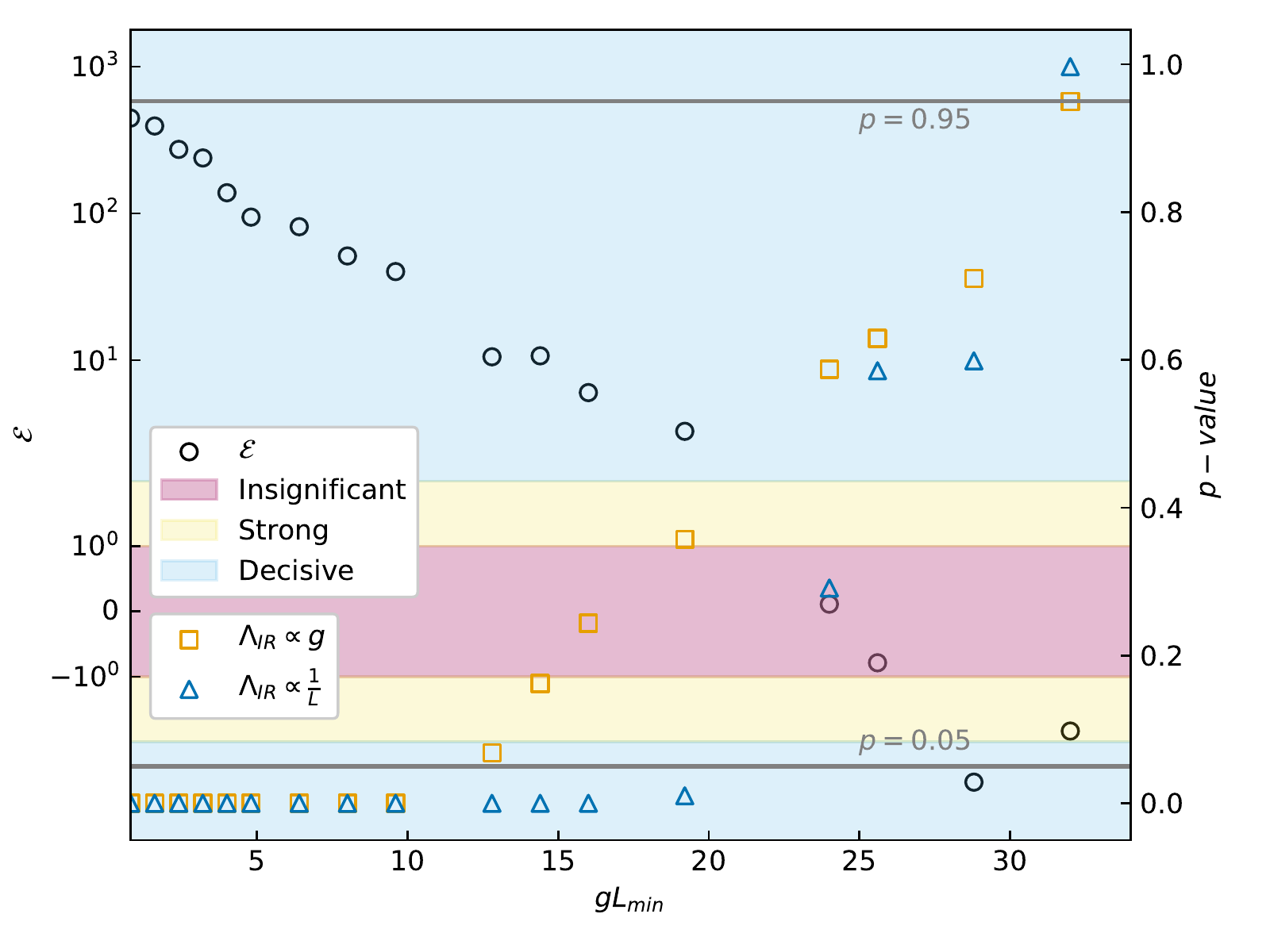}
    \caption{Top: $N = 2$, $\bar{B} = 0.52$, $\bar{B} = 0.53$ data, Bottom: $N = 4$, $\bar{B} = 0.42$, $\bar{B} = 0.43$ data. The $p$-value of the fit of equation (\ref{eq:msq fit}) with $\Lambda_{\rm IR} \propto g$ and $\Lambda_{\rm IR} \propto \frac{1}{L}$ (right $y$-axis) is shown by the orange squares and green triangles respectively. The black circles represent the log in base 10 of the Bayes Factor, $\mathcal{E} = \log_{10}(K)$, where $K = \frac{E_1}{E_2}$ with $E_1$ and $E_2$ being the marginal probabilities for fits with ($\Lambda_{\rm IR} \propto g)$ and ($\Lambda_{\rm IR} \propto 1/L$) respectively. The colored regions represent the strength of the evidence under the Jeffreys' Scale \cite{jeffreys1998theory}. Blue regions represent decisive evidence ($|\mathcal{E}| > 2$), while the yellow regions represent strong evidence ($1 < |\mathcal{E}| < 2$) and the pink region represents insignificant evidence ($|\mathcal{E}| < 1$).}
    \label{fig: pvalue-evidence}
\end{figure}

 The central fits are for pairs  $\Bbar=\{0.52,0.53\}|_{N=2}$ and $\{0.42,0.43\}|_{N=4}$, respectively, for which we found the largest number
 of degrees of freedom  described simultaneously.
The ansatz in Eq.~(\ref{eq:msq fit}) provides
an excellent parameterisation ($p$-values well above 5\%)
for the simulation data over the entire range $gL_{\rm min}\gtrsim 12$ to $gL_{\rm max}= 76.8$. The case $N=2$ is illustrated in Fig.~\ref{fig:centralfit} for $\Lambda_{\rm IR}=\frac 1{4\pi}\frac gN$. 
Tab.~\ref{tab:fitresults} summarises the fit results.  
The first error is statistical, and, where applicable, the second error is the maximum shift
of the fit result under variation of $gL_{\rm min}$ and the choice of $\Bbar$-pairs with
$\Bbar~\in~\{0.51,0.52,0.53,0.54,0.55,0.56,0.57,0.58,0.59\}|_{N=2}$ and
$\Bbar~\in~\{0.42,0.43,0.44,0.45,0.46,0.47\}|_{N=4}$, while requiring at least 
 15 degrees of freedom.
Note that the result for $\beta$ is compatible
with the prediction from perturbation theory, $\beta=1$ (cf. Eqs.~(\ref{eq:2loopI}) and~(\ref{eq:msq fit})). The result for $\nu$ for $N=2$ agrees well with a previous lattice determination~\cite{Hasenbusch:2000ph}, $\nu=0.710(2)$.
The EFT predictions for $\nu$ and $f(0)$ agree at the few-per-cent level (cf. Sec.~\ref{sec:EFT}).
Fits with $\Lambda_{\rm IR}\propto 1/L$ are not possible for similarly small values of $gL_{\rm min}$. For $N=2$ the first acceptable ($p\ge 0.05$) fit is possible only after discarding all data with $gL<32$ and for $N=4$, $gL<24$.
The r.h.s. axis in Fig.~\ref{fig: pvalue-evidence} shows how the $p$-value varies
with the cut in $gL$. 
Generally, larger $p$-values for $\Lambda_{\rm IR}\propto g$ at a
given value of $gL$ indicate that this ansatz provides a better description of the data
in terms of a $\chi^2$-analysis.
Inserting the the fit parameters in Tab.~\ref{tab:fitresults} into~(\ref{eq:msq fit}) and taking the limit $x\to\infty$ we obtain predictions for
the infinite-volume critical mass. For instance, for 
$ag=0.1$ we find $(am_c)^2=-0.031341(4)(6)$ for $N=2$ and $(am_c)^2=-0.045515(2)(7)$ for $N=4$. {If we assume IR power-divergences $D_{\rm IR}(x)\sim x^n$ ($n=1,2,3,4$) in lieu of logarithmic, no single acceptable fit was found ($p=0.00$).}
\begin{table*}
\begin{tabular}{llllllllllll}
\hline\hline
\multicolumn{1}{c}{$N$}&
\multicolumn{1}{c}{$gL_{\rm min}$}&
\multicolumn{1}{c}{$gL_{\rm max}$}&
\multicolumn{1}{c}{$\alpha_i$}&
\multicolumn{1}{c}{$\nu$}&
\multicolumn{1}{c}{$\beta$}&
\multicolumn{1}{c}{$f(0)$}&
\multicolumn{1}{c}{$f'(0)$}&
\multicolumn{1}{c}{$p$}&
\multicolumn{1}{c}{$\chi^2/N_{\rm dof}$}&
\multicolumn{1}{c}{$N_{\rm dof}$}
\\
\hline
  2&   12.8&76.8&0.0019(8)(18)&0.71(1)(6)& 1.05(5)(10)&0.577(1)(16)&-0.058(4)(53)& 0.2& 1.2& 31\\
\hline
  4&   12.8&76.8&\hspace{-1mm}\begin{tabular}{l}0.0010(5)(3)\\0.0014(4)(4)\end{tabular}&0.840(8)(8)&1.03(2)(2)&0.497(1)(5)& -0.090(3)(3)&0.07&1.4&30\\
  \hline\hline
\end{tabular}
\caption{Results of $\chi^2$ fits to finite-size-scaling data. The first error is statistical and the second systematic as described in the text.} \label{tab:fitresults}
\end{table*}

We also address the question of the IR regulator within the framework of Bayesian inference with uniform priors 
$\alpha \in [-0.4, 0.4]$,
$f(0) \in [0, 1]$, 
$f'(0) \in [-20, 20]$, 
$\beta \in [-15, 15]$ and
$\nu \in [0, 15]$. As in the frequentist study, for $N = 4$, two values of $\alpha$ are used: $\alpha_{1,2} \in [-0.4, 0.4]$. This analysis was done using Pymultinest \cite{Buchner:2014nha} as an interface to the MULTINEST \cite{feroz2008multimodal, Feroz:2008xx, feroz2013importance} code.
The marginalized probabilties of each model ($\Lambda_{\rm IR} \propto g$ and $\Lambda_{\rm IR} \propto 1/L$) were calculated across a range of $gL_{\rm min}$ cuts and pairings of $\bar{B}$ values. In Fig.~\ref{fig: pvalue-evidence} both the $p$-value, and the Bayes Factor of the central fit are shown across the range of $gL_{\rm min}$ values. In this plot, the graph is broken down into regions according to the Jeffreys' scale \cite{jeffreys1998theory}. The Bayes Factor $K$ is $\frac{E_1}{E_2}$, where $E_1$ and $E_2$ are the marginal probabilities for model 1 ($\Lambda_{\rm IR} \propto g)$ and model 2 ($\Lambda_{\rm IR} \propto 1/L$) respectively. If ${\cal E}=\log_{10}(K)$ is greater than $1$ there is strong evidence for model 1 over model 2, and if it is greater than $2$ it is decisive. The reverse is true for negative values of ${\cal E}$ in support of model 2. As the cut on $gL_{\rm min}$ is reduced (more data is used) the evidence for $\Lambda_{\rm IR} \propto  g$ increases, with there being  decisive evidence under the Jeffreys' scale for all $gL_{\rm min}$ cuts for $N = 2$ and for $gL_{\rm min} \leq 19.2$ cuts  for $N = 4$. The same pattern is seen for all $\bar{B}$ values.

One can also obtain parameter estimates via~the posterior probability distribution, which we find to be in excellent agreement with the results for the fit parameters from the $\chi^2$ analysis.

In conclusion, Bayesian inference prefers the IR-finite ansatz over the IR-divergent one; complementary and consistent with this, from $\chi^2$ fits we find the IR-finite FSS ansatz ($\Lambda_{\rm IR}\propto g$) able 
to describe more degrees of freedom (i.e.~larger range in $gL$) 
with acceptable $p$-value.

\paragraph{Conclusions and outlook:}
We present the first nonperturbative study of the critical properties of a three-dimensional super-renormalisable scalar QFT with $\varphi^4$ interaction and fields in the adjoint of $SU(N)$ with $N=2,4$. When studied in lattice perturbation theory, the theory exhibits
a logarithmic IR divergence for the critical mass at the two-loop level. The absence of this divergence in our numerical results from lattice simulations provides strong evidence for the IR-finiteness of the full theory. This constitutes one of the central results of this study.
Further results  are the nonperturbative determination of the 
    critical masses. For the range of couplings considered here the critical mass agrees 
    with 2-loop perturbation theory at and below the percent level when 
    employing the dimensionful coupling constant $g$ as IR regulator, 
    confirming the expectation of \cite{Jackiw:1980kv, Appelquist:1981vg}. Our result for the critical exponent is  close to the leading-order effective theory prediction, where the effective fields correspond to the magnetisation of the full theory.

Three-dimensional super-renormalisable QFT consisting of Yang-Mills theory coupled to adjoint scalar and/or fermionic matter are candidate theories for describing the physics of the early Universe by means of holographic duality. Our determination of the critical point constitutes the starting point towards the study of cosmology  from a three-dimensional QFT. In view of the holographic duality cosmic evolution corresponds to inverse RG flow where the initial singularity is mapped to the IR behaviour of the QFT. The absence of an
IR singularity on the QFT side may thus be seen as the holographic resolution of the initial singularity in the bulk.

\mbox{}\\

\begin{acknowledgments} The authors would like to warmly thank Pavlos Vranas for his valuable support during the early stages of this project. We would like to thank Masanori Hanada for collaboration at early stages for this project. A.J. and K.S. acknowledge funding from STFC consolidated grants ST/P000711/1 and ST/T000775/1.
A.P. is supported in part by UK STFC grant ST/P000630/1. A.P., J.K.L.L., and H.B.R are funded in part by the European Research Council (ERC) under the European Union’s Horizon 2020 research and innovation programme under grant agreement No 757646 and A.P. additionally grant agreement No 813942. J.K.L.L. is also partly funded by the Croucher foundation through the Croucher Scholarships for Doctoral Study. B.K.M. was supported by the EPSRC Centre for Doctoral Training in Next Generation Computational Modelling Grant No. EP/L015382/1. LDD is supported by an STFC Consolidated Grant, ST/P0000630/1, and a Royal Society Wolfson Research Merit Award, WM140078.
Simulations produced for this work were performed using the Grid Library (https://github.com/paboyle/Grid), which is free software under GPLv2. 
 This work was performed using the Cambridge Service for Data Driven Discovery (CSD3), part of which is operated by the University of Cambridge Research Computing on behalf of the STFC DiRAC HPC Facility (www.dirac.ac.uk). The DiRAC component of CSD3 was funded by BEIS capital funding via STFC capital grants ST/P002307/1 and ST/R002452/1 and STFC operations grant ST/R00689X/1. DiRAC is part of the National e-Infrastructure. The authors acknowledge the use of the IRIDIS High-Performance Computing Facility, and associated support services at the University of Southampton, in the completion of this work. 
 \end{acknowledgments}

\bibliography{IRreg}% Produces the bibliography via BibTeX.
%%%%%%%%%%%%%%%%%%%%%%%%%%%%%%%%%%%%%%%%%
\pagebreak
\widetext
\begin{center}
\textbf{\large Supplemental Materials: Nonperturbative IR finiteness in super-renormalisable scalar QFT}
\end{center}
%%%%%%%%%% Merge with supplemental materials %%%%%%%%%%
%%%%%%%%%% Prefix a "S" to all equations, figures, tables and reset the counter %%%%%%%%%%
\setcounter{equation}{0}
\setcounter{figure}{0}
\setcounter{table}{0}
\setcounter{page}{1}
\makeatletter
\renewcommand{\theequation}{S\arabic{equation}}
\renewcommand{\thefigure}{S\arabic{figure}}
\renewcommand{\bibnumfmt}[1]{[S#1]}
\renewcommand{\citenumfont}[1]{S#1}
%%%%%%%%%% Prefix a "S" to all equations, figures, tables and reset the counter %%%%%%%%%%

%%%%%%%%%%%%%%%%%%%%%%%%%%%%%%%%%%%%%%%%%%%%%%%%%%%%%%%%%%%%%%%%%%%%%%%
\section{IR behaviour of the two-loop correction}\label{sec:IR2loop}
%%%%%%%%%%%%%%%%%%%%%%%%%%%%%%%%%%%%%%%%%%%%%%%%%%%%%%%%%%%%%%%%%%%%%%%
The finite-size-scaling (FSS) analysis in this paper centers around the fit ansatz in Eq.~(\ref{eq:msq fit}), which uses the analytical expression for the 
IR-behaviour of the theory in Eq.~(\ref{eq:IR analytical}). 
We derive it by studying the IR behaviour of the 3$d$ lattice-regularised  2-loop integral,
\begin{equation}\label{sup:eq:2loopI}
D(p)=\int\limits_{-\pi/a}^{\pi/a}\frac {d^3k}{(2\pi)^{{3}}}\frac{d^{{3}}q}{(2\pi)^{{3}}}\frac 1{\hat k^2\,\hat q^2\, \hat r^2}\,,
\end{equation}
where $\hat k=\frac 2a \sin(ka/2)$, and $r=-k-q-p$.
The derivation is done conveniently in coordinate space, where
\begin{equation}\label{sup:eq:D(p)}
D(p)=\sum\limits_{x\in\Lambda}e^{-ipx}G(x)^3\,,
\end{equation}
with the coordinate-space scalar propagator $G(x)$.
Below we retrace the steps in $d=3$ taken by L\"uscher and Weisz~\cite{Luscher:1995zz} for $d=4$, to derive the large-$x$ expansion of the free scalar lattice propagator. 

The long-distance behaviour of the propagator should be independent of the discretisation. Following~\cite{Luscher:1995zz}, 
we therefore rewrite $G(x)$ in terms of the continuum scalar propagator and a smooth momentum cutoff, in this way restricting the integration to small momenta,
\begin{align}
G(x)=&\int\limits_{-\pi/a}^{\pi/a}\frac{d^3p}{(2\pi)^3} e^{ipx}\frac 1{\hat p^2}
\sim\int\limits_{-\infty}^{\infty}\frac{d^3p}{(2\pi)^3} e^{ipx}e^{-(ap)^2}\frac 1{  p^2},\nonumber\\
=&a^2 \int_1^\infty dt\int\limits_{-\infty}^{\infty}\frac{d^3p}{(2\pi)^3} e^{ipx}e^{-t(ap)^2}
=\frac 1{4\pi\sqrt{x^2}} {\rm Erf}\left[\frac{\sqrt{x^2}}{ 2 a}\right]\,.
\end{align}
For large $x$ we therefore expect
\begin{align}\label{eq:scalar prop largex}
G(x)\stackrel{x\to\infty}{\sim}&
\frac 1{4\pi}\frac 1{\sqrt{x^2}}\,.
\end{align}

We now introduce the auxiliary function,
\begin{equation}
H(x)=\int\limits_{-\pi/a}^{\pi/a}\frac{d^3p}{(2\pi)^3}e^{ipx}\ln\left((a\hat p)^2\right)\,,
\end{equation} 
and  the Vohwinkel relation~\cite{Vohwinkel},
\begin{equation}\label{eq:Vohwinkel}
\left(\delta^\ast_\mu+\delta_\mu\right)G(x)=x_\mu H(x)\,,
\end{equation}
with the lattice derivatives $\delta^\ast f(x)=\frac 1a (f(x)-f(x-\hat\mu a))$ and $\delta f(x)=\frac 1a (f(x+\hat\mu a)-f(x))$.

Eq.~(\ref{eq:Vohwinkel})  can be shown as follows: Consider the symmetrised lattice derivative
of the coordinate space propagator,
\begin{equation}
\left(\delta^\ast_\mu+\delta_\mu\right)G(x)=\frac 1a\int\limits_{-\pi/a}^{\pi/a} \frac{d^3p}{(2\pi)^3}\frac {2i\sin(ap_\mu)e^{ipx}}{\hat p^2}\,.
\end{equation}
Observing that 
\begin{align}
\frac{1}{a}\frac{2\sin ap_\mu}{(\hat p)^2}=\frac{\partial}{\partial p_\mu} \ln \left(\sum_{\nu} (a\hat p_\nu)^2\right)\,,
\end{align}
and using integration by parts we find
\begin{align}
\left(\delta^\ast_\mu+\delta_\mu\right)G(x) &= \frac 1a\int\limits_{-\pi/a}^{\pi/a}\frac{d^3p}{(2\pi)^3}\frac {2i\sin(ap_\mu)e^{ipx}}{\hat p^2}=
i\int\limits_{-\pi/a}^{\pi/a} \frac {d^3p}{(2\pi)^3}\left(\partial_{p_\mu} \ln \left((a\hat p)\right)^2\right)e^{ipx}\nonumber\\
&=x_\mu \int\limits_{-\pi/a}^{\pi/a}\frac{d^3p}{(2\pi)^3}\ln \left((a\hat p)\right)^2  e^{ipx} = x_{\mu} H(x)\,.
\end{align}

Using (\ref{eq:Vohwinkel}) for large $x$ we obtain
\begin{align}
H(x) \stackrel{x\to\infty}=
&\frac{1}{x_\mu}\partial_\mu G(x) =
-\frac 1{4\pi (x^2)^{3/2}}\,.
\end{align} 
Comparing with Eq.~(\ref{eq:scalar prop largex}) we  identify 
\begin{align}
G(x)^3\stackrel{x\to\infty}{=}&-\frac{1}{(4\pi)^2}H(x)\,,
\end{align}
which, by inverse Fourier transformation  allows us to conclude that
\begin{align}
D(p)\stackrel{p\to 0}{=}&-\frac {\ln((ap)^2)}{(4\pi)^2}\,.
\end{align}

%%%%%%%%%%%%%%%%%%%%%%%%%%%%%%%%%%%%%%%%%%%%%%%%%%%%%%%%%%%%%%%%%%%%%%%
\section{Finite-size scaling effective field theory}\label{sec:eft}
%%%%%%%%%%%%%%%%%%%%%%%%%%%%%%%%%%%%%%%%%%%%%%%%%%%%%%%%%%%%%%%%%%%%%%%

In this section we consider the continuum action~(\ref{eq:canonical S})
\begin{equation}
  S[\phi]=\frac{N}{g}\int\diff^3x\,\tr\left\{
  \sum_{\mu}[\partial_{\mu}\phi(x)]^2+(m^2-m_c^2)\phi(x)^2+\phi(x)^4
  \right\}\,,
  \label{eq:contthooftact}
\end{equation}
expressed as a function of the renormalised parameters $m^2$ and $g$.
%.............................................................................
\subsection{Effective theory}
The theory is expected to undergo a phase transition when the renormalised mass
becomes close to $0$ (\ie the correlation length diverges). In a finite cubic
volume, no transition can occur because no length in the system can exceed
the spatial extent, $L$. However, at the massless point, various statistical
quantities will scale non-trivially with $L$ according to the critical exponents
of the theory. Moreover, to analyse close-to-critical lattice simulation
results, it is important to understand the behaviour of finite-size effects.

In a periodic and cubic volume $\T^3$, a momentum vector $k$ is quantised as $\frac{2\pi}{L}n$,
where $n$ is a vector with integer components. In massless perturbation theory,
loop integrals become sums, such as
\begin{equation}
  I_1=\frac{1}{L^3}\sum_k\frac{1}{k^2}\,,
\end{equation}
for the tadpole diagram. Even with a UV regulator, such a sum is undefined because
of the explicit $\frac{1}{0}$ term it contains. This problem arises from a
sickness of the finite-volume free theory which is defined by a Gaussian
integral with a non-invertible covariance matrix. More explicitly, this matrix
is given by the Laplacian operator, which has an isolated zero-mode in the
finite-volume massless theory. However, in the full theory, the exponential in
the path integral is systematically damped by the quartic term $\tr[\phi(x)^4]$
in the action. This indicates that in the massless theory, the contribution from
the field zero-mode has to be treated non-perturbatively.

The magnetisation $M$ defined in~(\ref{eq:mag}) is the zero-momentum component of the field $\phi$ mentioned above, and we define the decomposition
\begin{equation}
  \phi=\chi+M\,,
\end{equation}
where $\chi$ has a vanishing zero-mode. Close to the critical regime, the theory will be dominated by the long-distance contributions from $M$. Therefore one can try to investigate finite-volume effects by using an effective theory where the higher-frequency modes $\chi$ are integrated out. We build this effective theory following the procedure described in~\citep[Sec. 37.3]{Zinn-Justin:572813}. The effective action $S_{\mathrm{eff}}[M]$ is defined by
\begin{equation}
  \label{eq:effaction}
  \exp(-S_{\mathrm{eff}}[M])=\frac{1}{C}\int\Diff\chi\,\exp(-S[\phi])\,,
\end{equation}
where $C$ is a normalisation factor defined by $S_{\mathrm{eff}}[0]=0$. The effective action has to be invariant under the $\mathbb{Z}_2$ symmetry $M\mapsto -M$ and the gauge symmetry $M\mapsto\Omega^{\dagger}M\Omega$ for any $\Omega$ in $\SU(N)$. This means that the only terms $S_{\mathrm{eff}}[M]$ can contain have the form $\tr(M^k)^l$ where $kl$ is an even integer.
%.............................................................................
\subsection{Leading-order effective action}
\label{sec:treeeft}
If one ignores entirely the corrections coming from the non-zero frequencies $\chi$, then it is clear from the original action~(\ref{eq:contthooftact}) that the effective action is given by
\begin{equation}
  \label{eq:treeeffaction}
  S_{\mathrm{eff}}[M]=\frac{L^3N}{g}[m^2\tr(M^2)+\tr(M^4)]\,,
\end{equation}
where the mass counter-term is absent because no dynamics from $\chi$ is included.
For an observable $O[M]$, the tree-level expression is given by
\begin{equation}
  \label{eq:treepathint}
  \braket{O[M]}=\frac{1}{\mathcal{Z}_{\mathrm{eff}}}
  \int_{\mathfrak{su}(N)}\diff M\,O[M]\exp(-S_{\mathrm{eff}}[M])\,,
\end{equation}
where $\mathcal{Z}_{\mathrm{eff}}$ is defined by $\braket{1}=1$. In other words, the effective theory is a random matrix theory on the space of traceless hermitian matrices. For an $\mathrm{SU}(N)$-invariant function $f$ on $\mathfrak{su}(N)$, the Weyl integration formula reduces integrating $f(M)$ over $\mathfrak{su}(N)$ to the integral over its $N-1$ independent eigenvalues
\begin{equation}
  \label{eq:weylint}
  \int_{\mathfrak{su}(N)}\diff M\,f(M)=
  \frac{\pi^{\frac{N(N-1)}{2}}}{\prod_{j=1}^N\Gamma(j)}
  \int\diff^{N-1}\lambda\,V(\bar{\lambda})^2f[\mathrm{diag}(\bar{\lambda})]\,.
\end{equation}
Here $\mathrm{diag}(\xi)$ is the diagonal matrix where the diagonal elements are the components of the vector $\xi$, ``barred'' vectors $\bar{\lambda}$ are defined by
\begin{equation}
  \bar{\lambda}=(\lambda_1,\dots,\lambda_{N-1},-{\textstyle\sum_{j=1}^{N-1}}\lambda_j)\,,
\end{equation}
and $V(\xi)$ is the Vandermonde determinant
\begin{equation}
  \label{eq:vdm}
  V(\xi)=\prod_{j<k}(\xi_j - \xi_k)\,,
\end{equation}
which is a homogenous polynomial of degree $\frac{1}{2}N(N-1)$. For future convenience, we additionally define
\begin{equation}
  \tilde{f}(\lambda)=f[\mathrm{diag}(\bar{\lambda})]\,.
\end{equation}
Using~(\ref{eq:weylint}), the expectation value of a gauge-invariant observable $O[M]$ in the tree-level effective theory is given by
\begin{equation}
  \braket{O[M]}=\frac{1}{\mathcal{Z}^\prime_{\mathrm{eff}}}
  \int\diff^{N-1}\lambda\,V(\bar{\lambda})^2\tilde{O}(\lambda)
  \exp\left\{-\frac{L^3N}{g}\left[m^2\bar{\lambda}^2
  +\bar{\lambda}^4\right]\right\}\,,
\end{equation}
where we use the notation $\xi^k=\sum_j\xi_j^k$ and where
$\mathcal{Z}^\prime_{\rm eff}$ is a trivial redefinition of $\mathcal{Z}_{\rm eff}$ to absorb the overall factor in~(\ref{eq:weylint}).
The most general gauge-invariant observable is given by $\tr(M^k)^l$. Using the change of variable $\lambda\mapsto(\frac{g}{L^3N})^{\frac{1}{4}}\mu$, one gets
\begin{equation}
  \label{eq:treemom}
  \braket{\tr(M^k)^l}=\left(\frac{g}{L^3N}\right)^{\frac{kl}{4}}
  \frac{\Psi_{kl}\left(m^2\sqrt{\frac{NL^3}{g}}\right)}
  {\Psi_{00}\left(m^2\sqrt{\frac{NL^3}{g}}\right)}\,,
\end{equation}
where the dimensionless function $\Psi_{kl}$ is defined by
\begin{equation}
  \label{eq:psikl}
  \Psi_{kl}(z)=\int\diff^{N-1}\mu\,
  V(\bar{\mu})^2(\bar{\mu}^k)^l
  \exp\left(-z\bar{\mu}^2
  -\bar{\mu}^4\right)\,,
\end{equation}
with the convention $0^0=1$. The Binder cumulant is then given by
\begin{equation}
  \label{eq:treebiinder}
  B=1-\frac{N}{3}\frac{\braket{\tr(M^4)}}{\braket{\tr(M^2)}^2}=f\left[\sqrt{N}\frac{m^2}{g^2}(gL)^{\frac32}\right]\,,
\end{equation}
with
\begin{equation}
  \label{eq:eftf}
  f(z)=1-\frac{N}{3}\frac{\Psi_{41}(z)\Psi_{00}(z)}{\Psi_{21}(z)^2}\,.
\end{equation}
Notice that at the critical point $m^2=0$, the Binder cumulant is independent from the volume and equal to $f(0)$. Moreover, this quantity is expected to be a function of $m^2(gL)^{1/\nu}$, which gives $\nu=\frac{2}{3}$ at this order of the effective expansion.

The Binder cumulant can also be written solely in terms of $\Psi_{00}(z)$. Indeed, from the definition (\ref{eq:psikl}) we immediately get 
\begin{equation}
    \Psi_{21}(z)= - \Psi'_{00}(z) \,,
\end{equation}
where prime indicate derivative w.r.t. $z$. Furthermore,
\begin{align}
\Psi_{41}(z) &= -\frac{\partial}{\partial w} \left[ \int\diff^{N-1}\mu\,
  V(\bar{\mu})^2
  \exp\left(-z\bar{\mu}^2
  -w \bar{\mu}^4\right)\right]_{w=1} \nonumber \\
&=-\frac{\partial}{\partial w} \left[w^{-\frac{1}{4} (N-1)(N+1)} \int\diff^{N-1}\mu\,
  V(\bar{\mu})^2
  \exp\left(-\frac{z}{w^{1/2}}\bar{\mu}^2
  - \bar{\mu}^4\right)\right]_{w=1} \nonumber \\
&=\frac{1}{4} (N-1)(N+1) \Psi_{00}(z) +\frac{1}{2} z \Psi'_{00}(z)\,,
\end{align}
where in the second line we rescaled the $\mu$'s such that the coefficient of $\bar{\mu}^4$ is equal to one.
It follows that 
\begin{equation} \label{eq:f_alternative}
f(z)=1-\frac{N}{3} \frac{  \frac{1}{4} (N-1)(N+1) +\frac{1}{2} z (\log \Psi_{00}(z))'}{((\log \Psi_{00}(z))')^2}   \,.
\end{equation}
\subsection{Determining the critical mass through the Binder cumulant}
At small $z$, the function $\Psi_{kl}(z)$ is a linear expansion of the form
\begin{equation}
  \Psi_{kl}(z)=\Psi_{kl}(0)- z\Psi_{kl}'(0)+\mathcal{O}(z^2)\,,
\end{equation}
where
\begin{equation}
  \label{eq:psiklprime}
  \Psi_{kl}'(0)=-\int\diff^{N-1}\mu\,
  V(\bar{\mu})^2(\bar{\mu}^k)^l\bar{\mu}^2
  \exp\left(-\bar{\mu}^4\right)\,.
\end{equation}
Using this expansion, one can expand the Binder cumulant function
\begin{equation}
  \label{eq:b4exp}
  f(z)=f(0)+zf'(0)+\mathcal{O}(z^2)\,,
\end{equation}
where $f'(0)$ is a combination of $\Psi_{kl}(0)$ and $\Psi_{kl}'(0)$ for $(k,l) \in [(4,1), (2,1), (0,0)]$. We now write the mass in terms of the bare and critical masses, $m^2=m_0^2-m_c^2$, where the $m_c^2$ is defined in the infinite-volume theory. Considering an arbitrary number $\bar{B}$ close to $f(0)$, we define the mass $\overline{m}^2(\bar B,g,L)$ to be the bare mass such that the Binder cumulant is equal to $\bar{B}$
\begin{equation}
  f\left\{\frac{\sqrt{N}}{g^2}\left[\overline{m}^2(L)-m_c^2\right](gL)^{\frac32}\right\}=\bar{B}\,.
\end{equation} 
Similar to Eq.~(\ref{eq:mc naive scaling}) we use the expansion~(\ref{eq:b4exp}) to obtain
\begin{equation}
  \overline{m}^2(L)=m_c^2+\frac{g^2}{\sqrt{N}}\frac{\bar{B}-f(0)}{f'(0)}
  \frac{1}{(gL)^{\frac32}}+\mathcal{O}\left(\frac{1}{L^3}\right)\,.
\end{equation}
The EFT predictions for $f(0)$ and $f'(0)$ can be evaluated numerically using standard  integration methods applied to the integrals~(\ref{eq:psikl}) and~(\ref{eq:psiklprime}). The values we found for $N=2,4$ are given in Tab.~\ref{tab:f0s}. 
\begin{table}[]
\begin{tabular}{l|l|l}
\hline\hline
$N$ & $f(0)$ & $f'(0)$  \\
\hline
2 &0.5431    &-0.03586\\
%3 &0.4341    &-0.01440\\
4 & 0.4459 &-0.02707 \\
\hline\hline
\end{tabular}
\caption{Numerical estimation from the EFT of the critical value of the Binder cumulant and the first derivative at the critical point. The function $f$ is defined in Eq.~(\ref{eq:eftf}).}
\label{tab:f0s}
\end{table}

\subsection{Asymptotics}

In this subsection we work out the asymptotic behavior of the Binder cumulant.
Eq.~(\ref{eq:f_alternative}) implies that all that we need is the asymptotic behavior of $\Psi_{00}$,
\begin{equation}
  \label{eq:psi00}
  \Psi_{00}(z)=\int\diff^{N-1}\mu\,
  V(\bar{\mu})^2
  \exp\left(-z\bar{\mu}^2
  -\bar{\mu}^4\right)\,.
\end{equation}
We start by discussing the case of general $N$ and then specialize to $N=2$ and $N=4$.

When $z>0$ we are in the symmetric phase, $\langle\bar{\mu}\rangle=0$, and in the large $z$ limit the quartic terms can be treated as a perturbation $\exp {\left(-\bar{\mu}^4\right)}\approx 1-\bar{\mu}^4$. At leading order ({\it i.e.} no $\bar{\mu}^4$),
one can obtain the parametric behavior of $\Psi_{00}$ by rescaling $\mu_i \to z^{-1/2} \mu_i$, leading to 
\begin{equation}
 \lim_{z \to \infty}   \Psi_{00}(z) \sim z^{-(N-1)(N+1)/2}\,,
\end{equation}
where we used the fact that $V(\bar{\mu})$ is a polynomial of degree $\frac{1}{2} N (N-1)$. Inserting in (\ref{eq:f_alternative}) we see the numerator of the second term on the r.h.s. vanishes and thus we need to include the first subleading correction, which can be easily done in generality since we are dealing with Gaussian integrals. We present the results for $N=2$ and $N=4$ below.

When $z<0$ we are in the broken phase and there is a non-trivial solution of the field equation. In the large $|z|$ limit, the $\bar{\mu}^2$ and  the $\bar{\mu}^4$ terms would need to balance each other and this implies that 
\begin{equation} \label{mus}
    \mu_i = \sqrt{|z|} \left(\mu_{(0)i} + \frac{\mu_{(1)i}}{|z|} + \cdots\right)\,.
\end{equation}
Then the leading behavior of $\Psi_{00}$ is 
\begin{equation} \label{broken}
 \lim_{z \to -\infty}   \Psi_{00} \approx |z|^{\frac{1}{2}N(N-1)- 2 k} e^{s z^2}, \qquad s = \bar{\mu}_{(0)}^2
  -\bar{\mu}_{(0)}^4 \; ,
\end{equation}
where $\bar{\mu}_{(0)} =(\mu_{(0)1}, \ldots, \mu_{(0) N-1},-(\mu_{(0)1} + \cdots \mu_{(0)N-1}))$ and $k$ is 
the number of coincident pairs of components in $\bar{\mu}_{(0)}$. This implies that 
\begin{equation}
\lim_{z \to -\infty} (\log \Psi_{00}(z))' \approx 2 s z\,,
\end{equation}
and therefore
\begin{equation}
\lim_{z \to -\infty} f(z) \approx 1- \frac{N}{12 s}\,.
\end{equation}

The $\mu_i$'s in (\ref{mus}) are determined by solving the field equations that follow from
\begin{equation} \label{eq:actionN}
    S(\mu_i; z) = S_0 + 2 \log V(\bar{\mu}), \qquad S_0 = -z\bar{\mu}^2
  -\bar{\mu}^4
\end{equation}
in the large $|z|$ limit. The leading order $\bar{\mu}_{(0)}$ follows from the solution of the field equations of 
$S_0$ and characterizes the symmetry breaking pattern of $SU(N)$. Given a solution $\bar{\mu}_{(0)}$ of the field equations of $S_0$ one needs to check whether this extends to a real perturbative solution of $S$.

When $N$ is even, which is the case of interest here,  there is a distinguished solution where all components are paired and have equal magnitudes
\begin{equation}
    \mu_{(0)i}^2 = \frac{1}{2}, \quad i=1, \ldots, N-1\, .
\end{equation}
When $N=2$ this is the only solution and when $N=4$ this is the only solution that extends as a real solution to subleading order in $|z|$. 
By inspection, for this solution $s=N/4$ and therefore universally 
\begin{equation} \label{eq:m2mininf}
\lim_{z \to -\infty} f(z) = \frac{2}{3}\, .
\end{equation}

\paragraph{$N=2$.}

Specializing (\ref{eq:psikl}) to $N=2$ yields,
\begin{equation} \label{eq:Psi00_N2}
    \Psi_{00}(z) = \int_{-\infty}^\infty  d\mu_1  e^{S[\mu_1;z]}\,,
\end{equation}
where
\begin{equation} \label{N2action}
    S(\mu_1;z)=-z (2 \mu_1^2) - 2 \mu_1^4 +  \log (2 \mu_1)^2 \,,
\end{equation}
where the logarithmic term is due to the Vandermonde determinant.

When $z>0$ we treat the quartic term as a perturbation:
\begin{equation}
\Psi_{00}(z) \approx  \int_{-\infty}^\infty  d\mu_1 (2 \mu_1)^2 e^{-z (2 \mu_1^2) } (1-  2 \mu_1^4) 
 = \sqrt{\frac{\pi}{2}} \frac{1}{z^{3/2}} \left(1-\frac{15}{8} \frac{1}{z^2} + O(\frac{1}{z^4})\right)\,.
\end{equation}
Inserting this in (\ref{eq:f_alternative}) and taking the $z \to \infty$ limit yields
\begin{equation}
    \lim_{z \to \infty} f(z) = \frac{4}{9} \approx 0.4444\, .
 \end{equation}

When $z<0$ we are in the broken phase and there is a non-trivial solution of the field equation of
(\ref{N2action}), 
\begin{equation}
    -|z| 2 \mu_1 + 4 \mu_1^3 - \frac{1}{\mu_1}=0 \, ,
\end{equation}
which in the $|z| \to \infty$ limit is given by
\begin{equation}
 \mu_1 = \pm \sqrt{|z|} \left(\frac{1}{\sqrt{2}} +\frac{1}{2 \sqrt{2} z^2}  + O(|z|^{-3})\right) \, .  
\end{equation}
Evaluating  (\ref{eq:Psi00_N2}) on this solution yields
\begin{equation}
    \Psi_{00}(z) \approx 2 z e^{\frac{z^2}{2}}\, , 
\end{equation}
which confirms (\ref{broken}) with $s=1/2$ and thus also (\ref{eq:m2mininf}).

\paragraph{$N=4$.} 
Specializing (\ref{eq:psikl}) to $N=4$ yields,
\begin{equation} \label{eq:Psi00}
    \Psi_{00}(z) = \int_{-\infty}^\infty \int_{-\infty}^\infty \int_{-\infty}^\infty d\mu_1 d\mu_2 d\mu_3 e^{S[\mu_i;z]}\,,
\end{equation}
where 
\begin{align}
 S[\mu_i;z]&=-z \left(\mu_1^2 + \mu_2^2 +\mu_3^2 + (\mu_1 + \mu_2 +\mu_3)^2\right) -
 \left(\mu_1^4 + \mu_2^4 +\mu_3^4 + (\mu_1 + \mu_2 +\mu_3)^4\right) \nonumber \\
 &\quad +2 \log\left((\mu_1-\mu_2) (\mu_1-\mu_3) (\mu_2-\mu_3) (2 \mu_1+\mu_2 + \mu_3)( \mu_1+2 \mu_2 + \mu_3) (\mu_1+\mu_2 + 2 \mu_3) \right)\, , 
\end{align}
where the second line is due to the Vandermonde determinant.

When $z>0$ we are in the symmetric phase and treating the quartic terms as a perturbation we obtain, 
\begin{equation}
    \Psi_{00}(z) \approx \frac{9 \pi^{3/2}}{4 z^{15/2}}\left(1-\frac{435}{16 z^2} + O(\frac{1}{z^4}))\right) \, .
 \end{equation}
Inserting this in (\ref{eq:f_alternative}) and taking the $z \to \infty$ limit yields
\begin{equation}
    \lim_{z \to \infty} f(z) = \frac{16}{45} \approx 0.3556\, .
 \end{equation}

When $z<0$ we are in the broken phase and the field equations admit the following solution for large $|z|$,
\begin{equation}
    \mu_1 = \sqrt{|z|} \left(-\frac{1}{\sqrt{2}} -\frac{1}{2 |z|} + O(\frac{1}{|z|^2})\right), \quad 
     \mu_2 = \sqrt{|z|} \left(\frac{1}{\sqrt{2}} -\frac{1}{2 |z|} + O(\frac{1}{|z|^2})\right), \quad
      \mu_3 = \sqrt{|z|} \left(-\frac{1}{\sqrt{2}} +\frac{1}{2 |z|} + O(\frac{1}{|z|^2})\right)\, .
\end{equation}
Evaluating  (\ref{eq:Psi00}) for this saddle yields
\begin{equation}
    \Psi_{00}(z) \approx 16 z^2 e^{z^2}\, , 
\end{equation}
which confirms (\ref{broken}) with $s=1$ and thus also (\ref{eq:m2mininf}) (note that $k=2$ in this case since $\bar{\mu}_{(0)4} = \frac{1}{\sqrt{2}}$) .

\subsection{Exact formulas for $N=2$}
For $N=2$,~(\ref{eq:psikl}) is a one-dimensional integral that can be computed explicitly
\begin{align}
  \label{eq:treepsiklsu2}
  \Psi_{kl}(z) &= e^{-\frac{1}{2}z^2} \left({\rm sign}(z)\right)^{\frac{1}{2} (3 +k l)} 2^{\frac{1}{4} 
  -\frac{1}{4} k l + l} I_{kl} \\
   I_{kl}(z) & =\Gamma \left(\frac{kl+3}{4}\right)\, _1F_1\left(-\frac{kl+1}{4};\frac{1}{2};-\frac{z^2}{2}\right)-\sqrt{2} z \Gamma
   \left(\frac{kl+5}{4}\right) \, _1F_1\left(\frac{1-k l}{4};\frac{3}{2};-\frac{z^2}{2}\right)\,,
\end{align}
where $\,_1F_1$ is the confluent hypergeometric function.

Substituting in (\ref{eq:eftf}) we find
\begin{equation}
  \label{eq:eftfN2}
  f(z)=1-\frac{1}{3}\frac{I_{41}(z) I_{00} (z) }{(I_{21}(z))^2}\,.
\end{equation}
The zero and the asymptotes can be obtained using well-known properties of confluent hypergeometric functions and are given by  
\begin{equation}
    f(0)=1-\frac{\Gamma(3/4) \Gamma(7/4)}{3 \Gamma(5/4)^2} \approx 0.5431, \qquad \lim_{z \to -\infty} f(z) = \frac{2}{3}, \qquad  \lim_{z \to \infty} f(z) = \frac{4}{9}.
\end{equation}
The asympotic values confirm the values we obtained in the previous subsection.

\section{Bayesian Analysis}
The primary quantity used here to determine the favoured form of the infrared regulator, $\Lambda_{\rm IR}$, is the Bayes Factor, $K$. This quantity is equal to the ratio of the marginalised probabilities of one model over the other, given the data. We can derive its expression through application of Bayes Theorem.

\begin{align}
    p(M|\mathrm{data}) &= \int d\alpha\,  p(M (\alpha)|\mathrm{data})\, p(\alpha|M), \label{eq: Bayes Theorum}\\
              &= \int d\alpha\, \frac{p(\mathrm{data}| M(\alpha))\, p(M(\alpha))}{p(\mathrm{data})}\, p(\alpha|M), \nonumber\\
              &= \frac{p(M)}{p(\mathrm{data})}\, \int d\alpha\, L(M(\alpha))\, p(\alpha|M), \nonumber
\end{align}
where $M$ is a model of the data, $\alpha$ is the parameters of that model. The first line of (\ref{eq: Bayes Theorum}) simply separates the probability of a model given the data into the contributions from all possible parameter values, weighted by the prior of those parameter values, $p(\alpha|M)$. In the second line Bayes theorem is applied. In the final line the definition of the likelihood function, $L$ is used. Taking the ratio of (\ref{eq: Bayes Theorum}) between two competing models yields the Bayes Factor,

\begin{align}
    K(\mathrm{data}) = \frac{p(M_1|\mathrm{data})}{p(M_2|\mathrm{data})} = \frac{\int d\alpha_1\, L(M(\alpha_1))\, p(\alpha_1|M)}{\int d\alpha_2\, L(M(\alpha_2))\, p(\alpha_2|M)}\; \frac{p(M_1)}{p(M_2)}. \label{eq: Likelihood ratio}
\end{align}

The fraction $\frac{p(M_1)}{p(M_2)}$ expresses any prior belief of the likelihood of one model over the other. In this Letter, as is often the case, this has been set to $1$. In this analysis, $M_1$ and $M_2$ are both of the form given in equation (\ref{eq:msq fit}), where they differ only by the expression of $\Lambda_{\rm IR}$, with model 1 using $\Lambda_{\rm IR} = \frac{1}{4 \pi} \frac{g}{N}$ and model 2 using $\Lambda_{\rm IR} = \frac{1}{L}$. The models therefore share the same model parameters, and the same uniform priors. Under the statistical bootstrap \cite{Efron}, the probability density of the data points is assumed to follow a multivariate Gaussian distribution, with covariance matrix, $\mathrm{Cov}$, estimated from the bootstrap samples. Writing the data as a vector of inputs $\textbf{x}$, and output masses $\textbf{m}$, we have the familiar Gaussian distribution for $L(M(\alpha))$:

\begin{align}
    L(M(\alpha)) = \frac{1}{\sqrt{(2 \pi) ^ k |\mathrm{Cov}|}} \exp\left( - \frac{1}{2} (\textbf{m} - M(\alpha))(\textbf{x}))^T \cdot \mathrm{Cov}^{-1} \cdot (\textbf{m} - M(\alpha))(\textbf{x}))\right)\, ,
\end{align}
where we have denoted the dimension of the data vector $\textbf{m}$ by $k$.

Since the covariance matrix is defined through the bootstrap independently of the parameters $\alpha$, the pre-factor to the exponential may be brought outside of the integral, where it cancels between the numerator and denominator of $K$. Taking a logarithm of $K$ gives the log of the Bayes factor, which one can interpret using the Jeffreys' scale \cite{jeffreys1998theory}.\\

\begin{figure}
    \centering
    \includegraphics[width=160mm]{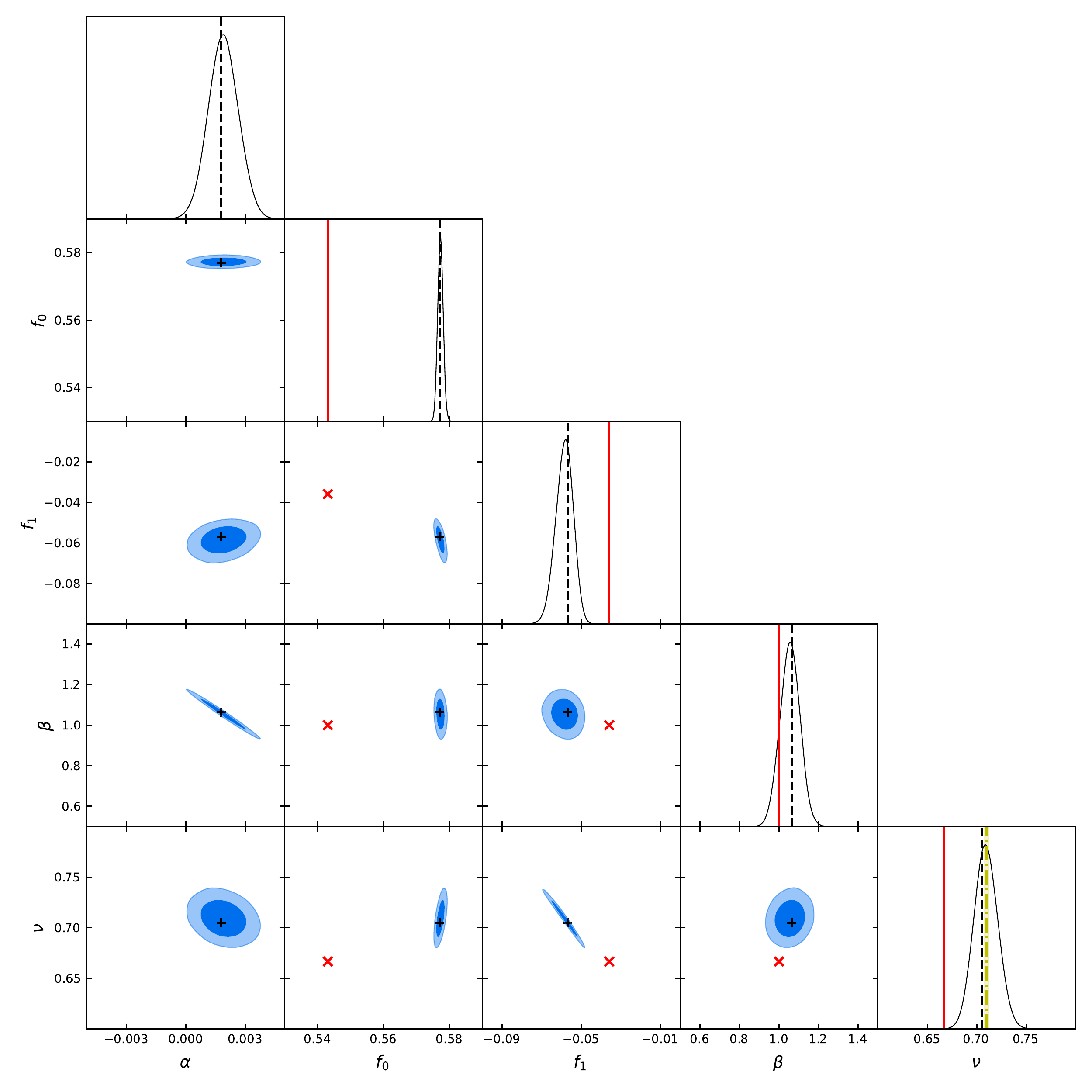}
    \caption{Posterior probability density obtained using \cite{Feroz:2008xx, feroz2008multimodal, feroz2013importance, Buchner:2014nha} and plotted with \cite{Lewis:2019xzd}, for $N = 2$ data with $\bar{B} = 0.52$ and $\bar{B} = 0.53$ and a $gL_{min}$ cut of $12.8$. The red (`x') points and red-solid lines are the predictions of the EFT. The black (`+') points and the black-dashed lines show the parameters of the maximum likelihood estimate. The yellow dot-dashed line shows the value of $\nu$ found in \cite{Hasenbusch:2000ph}.}
    \label{fig: posterior}
\end{figure}

\begin{figure}
    \centering
    \includegraphics[width=190mm]{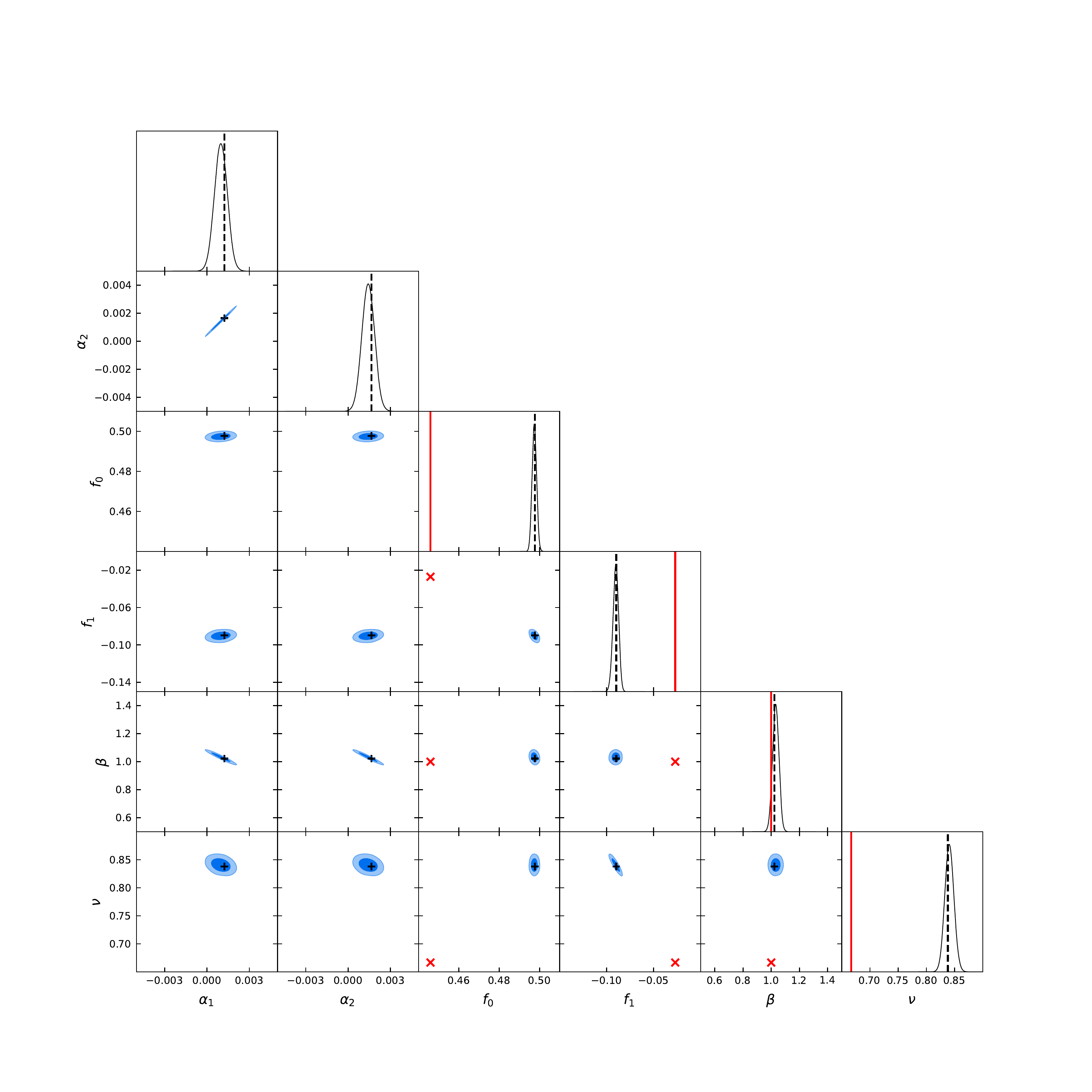}
    \caption{Posterior probability density obtained using \cite{Feroz:2008xx, feroz2008multimodal, feroz2013importance, Buchner:2014nha} and plotted with \cite{Lewis:2019xzd}, for $N = 4$ data with $\bar{B} = 0.42$ and $\bar{B} = 0.43$ and a $gL_{min}$ cut of $12.8$. The red (`x') points and red-solid lines are the predictions of the EFT. The black (`+') points and the black-dashed lines show the parameters of the maximum likelihood estimate.}
    \label{fig: posterior4}
\end{figure}

The integrand of the Bayesian evidence integral, $L(M(\alpha))\, p(\alpha|M)$, is significant as it is the posterior probability distribution of the model parameters given the data. As an example, the posterior distribution of the model parameters for the central fits of $N=2$ and $N=4$ are shown in figures \ref{fig: posterior} and \ref{fig: posterior4} respectively.

\newpage
\bibliography{IRreg}

%%%%%%%%%%%%%%%%%%%%%%%%%%%%%%%%%%%%%%%%%
\end{document}